\documentclass[pdflatex,sn-nature]{sn-jnl}

\usepackage{graphicx}%
\usepackage{multirow}%
\usepackage{amsmath,amssymb,amsfonts}%
\usepackage{amsthm}%
\usepackage[cal=cm, scr=rsfso]{mathalfa}%
\usepackage[title]{appendix}%
\usepackage{xcolor}%
\usepackage{textcomp}%
\usepackage{manyfoot}%
\usepackage{booktabs}%
\usepackage{algorithm}%
\usepackage{algorithmicx}%
\usepackage{algpseudocode}%
\usepackage{listings}%
\usepackage{float}%

\makeatletter
\def\addressfont{\small\titraggedcenter}
\makeatother

\begin{document}

\title[Rapid and Cost-Effective In situ Fabrication of Nanoporous Membranes in Microfluidic Devices for Biomedical and Environmental Applications]{Rapid and Cost-Effective In situ Fabrication of Nanoporous Membranes in Microfluidic Devices for Biomedical and Environmental Applications}

\author[1]{\fnm{Arijit} \sur{Mohanta}}

\author[1]{\fnm{Zakia} \sur{Farhat}}

\author[1]{\fnm{Yeshudan} \sur{Bora}}

\author[2]{\fnm{Saurabh} \sur{Dubey}}

\author[3]{\fnm{Nafisa} \sur{Arfa}}

\author*[1,2,4]{\fnm{Dipankar} \sur{Bandyopadhyay}}

\affil[1]{\orgdiv{Department of Chemical Engineering}, \orgname{Indian Institute of Technology Guwahati}, \orgaddress{\city{Guwahati}, \state{Assam}, \postcode{781039}, \country{India}}}

\affil[2]{\orgdiv{Centre for Nanotechnology}, \orgname{Indian Institute of Technology Guwahati}, \orgaddress{\city{Guwahati}, \state{Assam}, \postcode{781039}, \country{India}}}

\affil[3]{\orgdiv{Davidson School of Chemical Engineering}, \orgname{Purdue University}, \orgaddress{\city{West Lafayette}, \state{IN}, \postcode{47907}, \country{USA}}}

\affil[4]{\orgdiv{Jyoti and Bhupat Mehta School of Health Sciences and Technology}, \orgname{Indian Institute of Technology Guwahati}, \orgaddress{\city{Guwahati}, \state{Assam}, \postcode{781039}, \country{India}}}

\abstract{Micro/nanoporous membranes have been used extensively in biological and medical applications associated with filtration, particle sorting, cell separation, and real-time sensing. Expanding its applications to microfluidics offers several advantages, such as precise control over fluid flow rates, regulated reaction times, and efficient product extraction. This study demonstrates a simple, cost-effective, rapid, single-step method for the synthesis of nanoporous membranes within a microfluidic device. The membrane is prepared by filling up the microchamber with cellulose acetate dissolved in N, N-dimethylformamide, and 1-hexanol, followed by the continuous flow of water as an antisolvent. Mixing CA-DMF with antisolvent water leads to an in-situ synthesis of cellulose acetate nanoparticles (CANPs) at the Water-DMF interface. The continuous flow of water ensures the coagulation of CANPs, forming nanoporous membranes. The thickness, porosity, and wettability of the membrane are dependent on the flow rate of water and the concentration of CA dissolved in DMF. Apart from its wide application in cell and biomolecule separation, this membrane can be used to detect biomarkers or pathogens in clinical samples. Additionally, these nanoporous membranes can be used to detect and filter environmental contaminants such as heavy metals, pesticides, and emulsified oil from water. In summary, the synthesis of CANP nanoporous membranes within microfluidic channels with adjustable porosity enhances the potential uses of micro/nanomembranes in both healthcare and environmental contexts.}


\maketitle

\section{Introduction}\label{sec1}

Porous membranes have versatile uses in industry \cite{Buonomenna2013}, healthcare applications \cite{Sun2021}, agriculture \cite{Ma2021}, and wastewater treatments \cite{Quist-Jensen2015} due to their large surface area \cite{Holst2010}, adjustable pore sizes \cite{Lin2021}, and strong mechanical properties \cite{Lv2017}. As the demand for enhanced filtration, separation, and sensing technologies intensifies, the development of micro/nanoporous membranes has emerged as a crucial advancement in domains such as medical diagnostics \cite{Adiga2009} and environmental technologies \cite{Qi2015}.\\

In biomedical applications, micro/nanoporous membranes have assumed prime significance, especially for creating in vitro cell coculture systems that mimic in vivo microenvironments to study cell-cell interactions \cite{Casillo2017}, drug responses \cite{Arruebo2012}, and tissue engineering \cite{Hadjizadeh2010}. Alongside, these membranes are also used in medical equipment for purposes such as dialysis, where they selectively remove waste elements from the blood while leaving important proteins and cells intact \cite{Himmelfarb2020, Lin2016, Ferraz2013}. Furthermore, their characteristic selective permeability imparts superior accuracy when integrated with biosensors to detect varied biomarkers \cite{DeLaEscosura-Muiz2011, Vinoth2023, Arshavsky-Graham2021, Westphalen2020}. This enhancement considerably improves the sensors' sensitivity and precision \cite{Zhou2019, Sola2015, Schuller2020, Reimhult2008, Ivanauskas2008}. Additionally, porous membranes have been used in environmental applications to detect and filter contaminants, including heavy metals \cite{Huang2014, Sulaiman2020}, pesticides \cite{Abdelhameed2021, Goh2022, Plakas2012}, and emulsified oils \cite{Peng2016, Zhu2014} from water, which helps to monitor and preserve the environment. Their ability to precisely control filtration and separation processes makes them indispensable in addressing environmental pollution and ensuring clean water and air \cite{Baig2020, Ma2019, Dharupaneedi2019}.\\

Despite their multifarious advantages, traditional fabrication methods for micro/nanoporous membranes often involve complex, time-consuming, and costly procedures. These conventional techniques involve multiple steps, requiring sophisticated equipment and specialized knowledge, making it difficult to scale up production for practical use \cite{Guo2022, Shiohara2021, Stucki2018, Lalia2013, Li2021, Ionita2016}. In light of this background, there is a pressing need for simpler, more cost-effective methods that can produce high-quality nanoporous membranes without compromising their functional properties.\\

Addressing these challenges, recent advances in research have introduced a rapid, single-step method for implementing nanoporous membranes within microfluidic devices. The integration of these membranes into microfluidic systems offers several advantages, including precise control over fluid flow rates, regulated reaction times, and efficient product extraction \cite{Zhu2017, Chen2016, Lechl2015, DeJong2006, Wei2011}. However, incorporating a micro/nanoporous membrane inside a microchannel remains challenging due to the need for precise accuracy, fabrication complexity, and material compatibility issues. In this paper, we have demonstrated an in-situ fabrication approach for creating nanoporous membranes from the coagulation of cellulose acetate nanoparticles within a microchannel. The physical properties of these membranes have been characterized and explored in terms of their potential applications.\\

In conventional methods, membranes fabricated using lithography are inserted into microchannels to achieve filtration. While these membranes often demonstrate higher filtration rates in each cycle, the process is complicated and expensive, requiring sophisticated equipment and skilled operators. In contrast, our technique involves the in-situ fabrication of nanoporous membranes directly within the microchannel. This approach is not only simpler and more cost-effective but also allows for easy adjustment of pore density to suit specific filtration needs. In our experiment, discussed below, a filtration efficiency of 57.52\% for methyl red dye was achieved at the end of the fourth cycle. While achieving filtration rates comparable to conventional methods may require additional filtration cycles, our process offers distinct advantages, including faster operation, reduced costs, portability, and ease of use. These features make our technique a promising alternative to traditional methods for efficient dye and microparticle filtration in microfluidic applications.\\

\section{Experimental Section}\label{sec2}

\subsection{Materials}\label{subsec1}
Cellulose acetate (CA) powder, N,N-dimethylformamide (DMF), and 1-Hexane were procured from Sigma-Aldrich. Green fluorescent-tagged polystyrene beads (2~$\mu$m) and methyl red dye were also purchased from Sigma-Aldrich. Polydimethylsiloxane (PDMS) and the crosslinker Sylgard-128 were obtained from Dow Corning, India. Acrylonitrile Butadiene Styrene (ABS) filament and AA 0.25 print core for the Ultimaker-3 were supplied by Imaginarium Solutions, Mumbai, India. All chemicals utilized were of analytical grade. Milli-Q water was used for all experimental purposes.

\subsection{Methods}\label{subsec2}
\subsubsection{Solution Preparation}\label{subsubsec1}
The cellulose acetate-containing solution is prepared by dissolving cellulose acetate powder in N, N-dimethylformamide (DMF), and 1-hexanol. Initially, 500~mg CA and 5 ml of DMF are combined in a centrifuge tube. After that, 5 ml of 1-hexanol is added to the mixture. The centrifuge tube is shaken thoroughly using a vortex shaker until all of the CA is completely dissolved in the mixture. Deionized water is used as the antisolvent for the formation of CANPs.

\subsubsection{Micochannel Fabricaion}\label{subsubsec2}
\begin{figure}[H]
    \centering
    \includegraphics[width=1\textwidth]{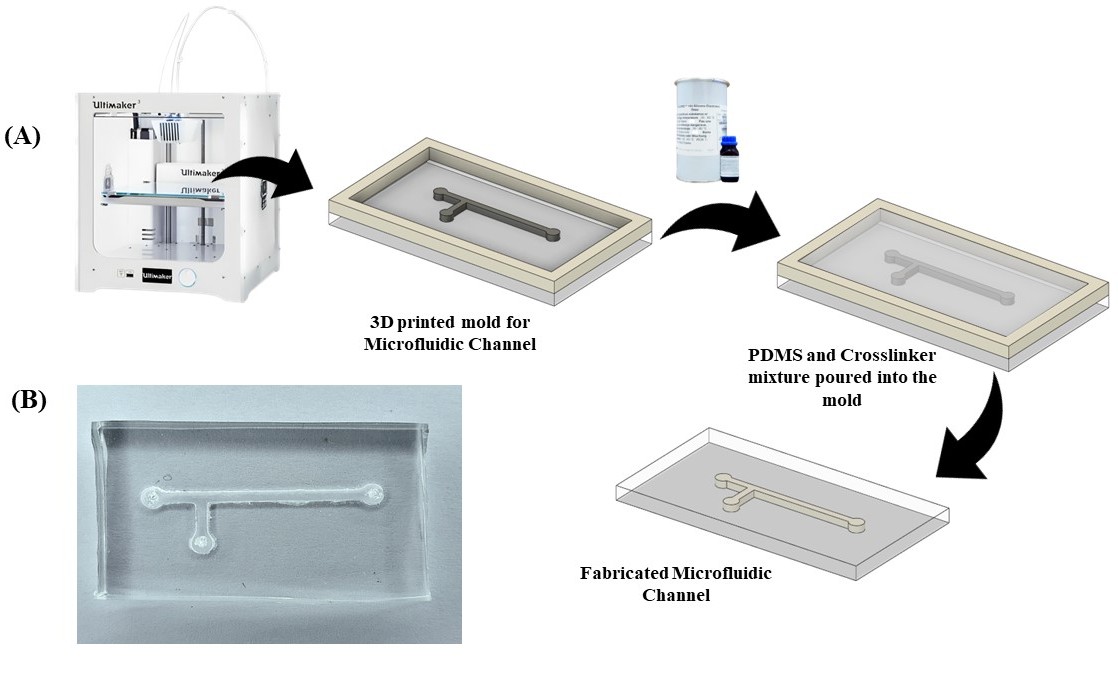} 
    \caption{(A) Schematic diagram illustrating the fabrication process of the microfluidic channel. (B) Fabricated T-shaped microfluidic channel.}
    \label{Figure 1}
\end{figure}

A microfluidic channel is fabricated following the replica molding soft-lithography technique. Initially, the mold and trench for the T-shaped microchannel are 3D printed using an Ultimaker-3 3D printer equipped with an AA 0.25 print core. Acrylonitrile Butadiene Styrene (ABS) is used as the 3D printing material. The mold, along with the trench, is attached to a glass plate. Afterward, a 10:1 v/v mixture of PDMS and Sylgard 184 is poured inside the trench and cured at 90 °C for 2 hours. Following curing, the hardened PDMS structure is bonded to a glass slide using the silanization method, forming the microchannel (Fig. \ref{Figure 1}).

\subsubsection{Membrane Synthesis inside Microchannel}\label{subsubsec3}
\begin{figure}[t!]
    \centering
    \includegraphics[width=0.9\textwidth]{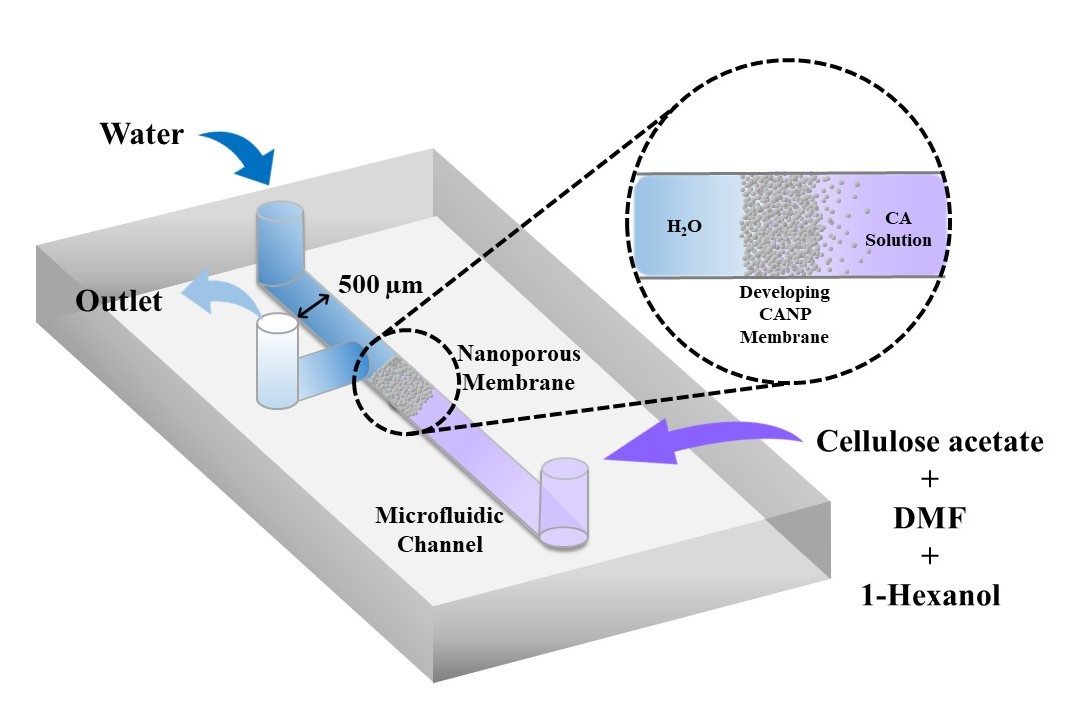} 
    \caption{Fabrication of cellulose acetate membrane within a microfluidic channel, occurring at the interface of water and a mixture of cellulose acetate, DMF, and 1-hexanol.}
    \label{Figure 2}
\end{figure}

The membrane is prepared by filling the microchamber with CA dissolved in DMF and 1-hexanol, followed by the continuous flow of water as an antisolvent. At first, the microchannel is filled with the CA solution all the way to the T-shaped joint. Subsequently, the water flow is started from the opposite inlet. An annular path is used to drain out the air trapped inside the microchannel. Upon contact between the water and the CA solution, the annular path was closed to prevent water bypass. The water flow rate is maintained at 0.001~ml/min. Mixing CA solution with antisolvent water results in an in-situ fabrication of cellulose acetate nanoparticles (CANPs) at the interface of both solutions (Fig. \ref{Figure 2}).\\

\begin{figure}[b!]
    \centering
    \includegraphics[width=0.8\textwidth]{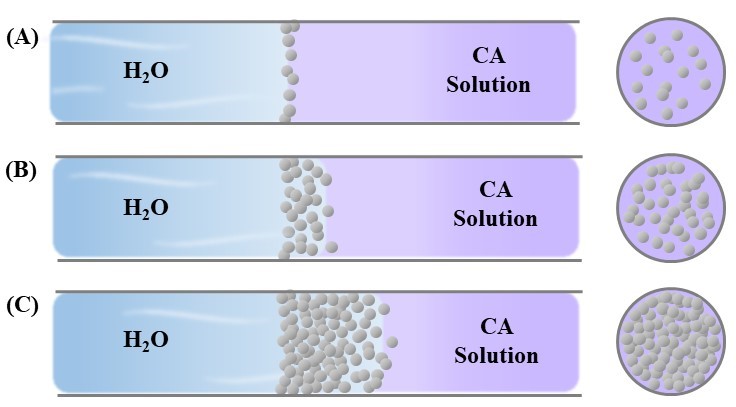} 
    \caption{Longitudinal and transverse cross-sectional representation of fabrication of cellulose acetate membrane within the microfluidic channel, with continuous flow of antisolvent water. As DMF diffuses into the water, cellulose acetate nanoparticles (CANPs) continuously form and coagulate.}
    \label{Figure 3}
\end{figure}

The formation of CANPs can be understood through the interaction between DMF, cellulose acetate, and water. DMF exhibits a strong compatibility for water, whereas CA is insoluble in water. When DMF and water come into contact, DMF rapidly diffuses into the water phase. This diffusion results in the phase separation of CA particles at the solution interface, resulting in the creation of CANPs. Prolonged interaction between DMF and water results in substantial CANP precipitation at the interface (Fig. \ref{Figure 3}). The continuous flow of water ensures the coagulation of CANPs, thereby forming nanoporous membranes (Fig. \ref{Figure 4}A).\\

\section{Results \& Discussions}\label{sec3}
\subsection{Characterization of Membrane}

\begin{figure}[b!]
    \centering
    \includegraphics[width=0.8\textwidth]{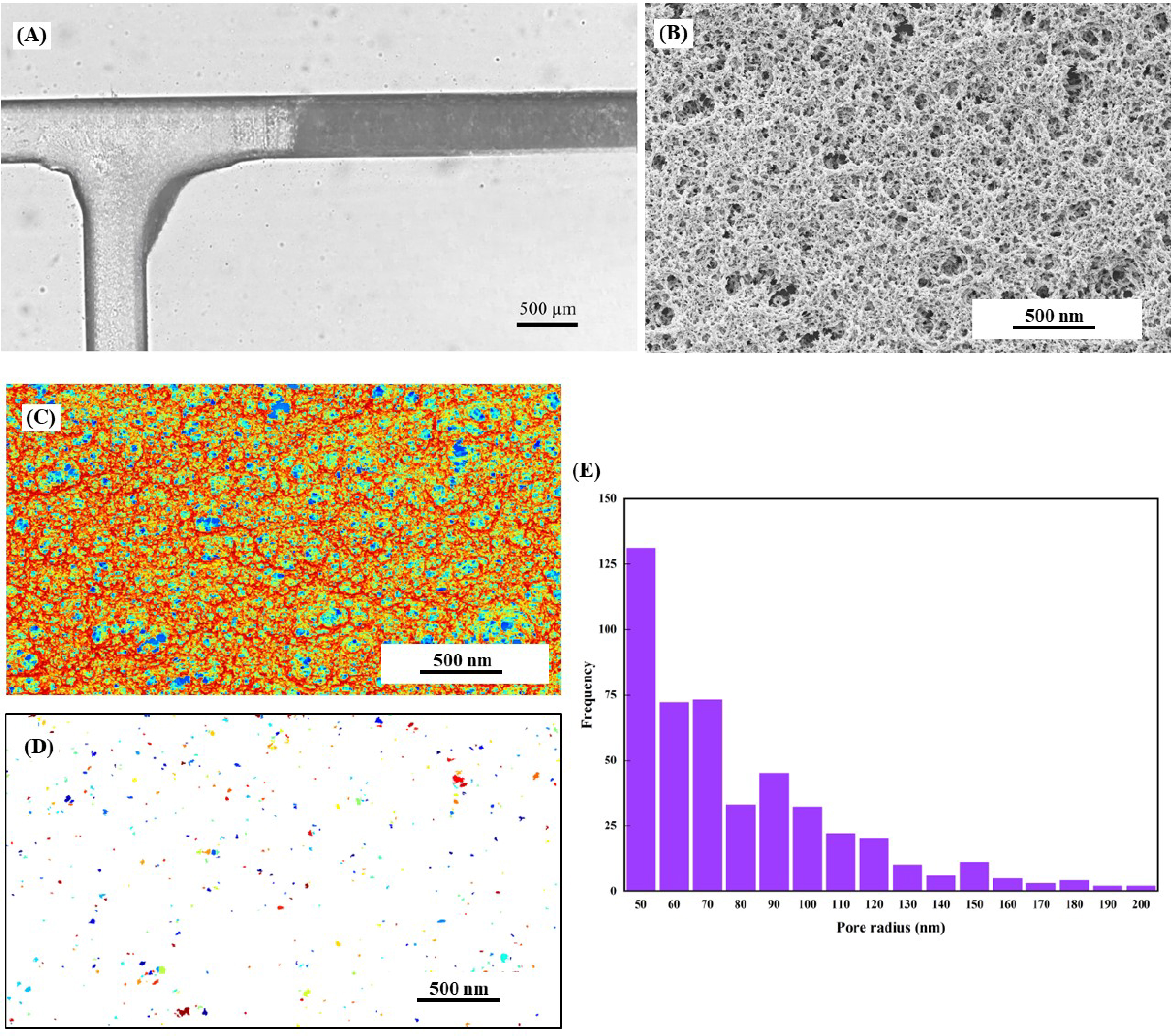} 
    \caption{(A) Nanoporous cellulose acetate membrane fabricated inside the microfluidic channel. (B) FESEM image of the nanoporous membrane. (C) Depth map of the membrane. (D) Pore space segmentation of the membrane. (E) Pore size distribution plot. Images (C), (D), and bar plots (E) were obtained by performing image analysis of the image (B).}
    \label{Figure 4}
\end{figure}

The morphology of the cellulose acetate (CA) membrane was studied using FESEM. To prepare for FESEM analysis, two parallel transverse cuts with a very tiny gap were made through the membrane, which was constructed within the microchannel. The produced samples were then dried under vacuum overnight to remove any remaining solvents, resulting in a clearer and more precise image of the membrane structure. A thin platinum layer was deposited before FESEM imaging. The pore size of the membrane was also evaluated via FESEM analysis (Fig. \ref{Figure 4}B). The MATLAB image analysis (Fig. \ref{Figure 4}C-E) reveals that the fabricated nanoporous membrane has pore diameters ranging from approximately 100 to 200~nm with an average pore diameter of 160~nm and porosity of 0.0194.

\subsection{Dye Separation}
\begin{figure}[H]
    \centering
    \includegraphics[width=0.9\textwidth]{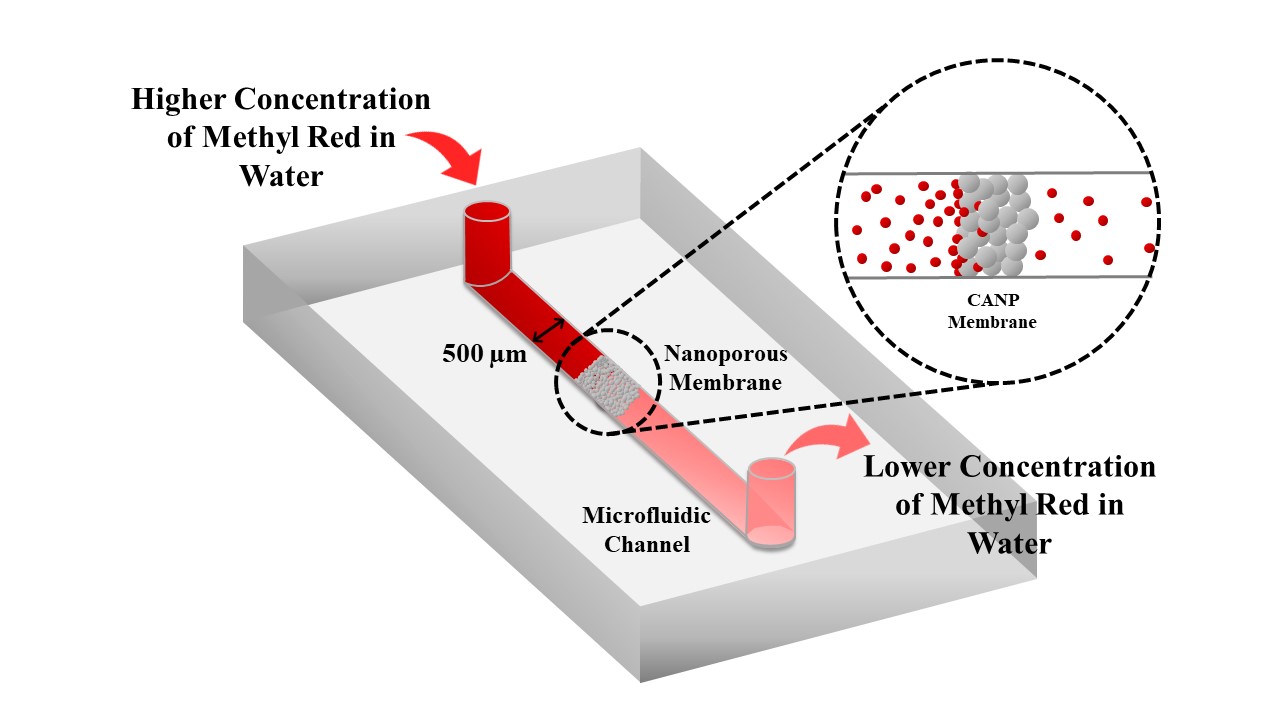} 
    \caption{Filtration of methyl red dye using cellulose acetate nanoporous membrane within microchannel. The inlet introduces a high-concentration methyl red solution, which is filtered through the membrane, resulting in a lower-concentration permeate exiting from the outlet.}
    \label{Figure 5}
\end{figure}

In this experiment, the T-shaped channel’s water inlet and the CA solution inlet are used as the inlet and outlet for the methyl red solution, respectively. To prevent bypass flow, the annular channel is closed. A 1~mM solution of methyl red dye dissolved in water is inserted into the system and flowed throughout the membrane at a controlled flow rate of 0.1~ml/min (Fig. \ref{Figure 5}). It is observed that the nanoporous CA membrane is able to effectively facilitate dye separation, with the retentate holding the concentrated dye solution and the permeate showing a substantially lower quantity of methyl red.\\

\begin{figure}[t!]
    \centering
    \includegraphics[width=0.9\textwidth]{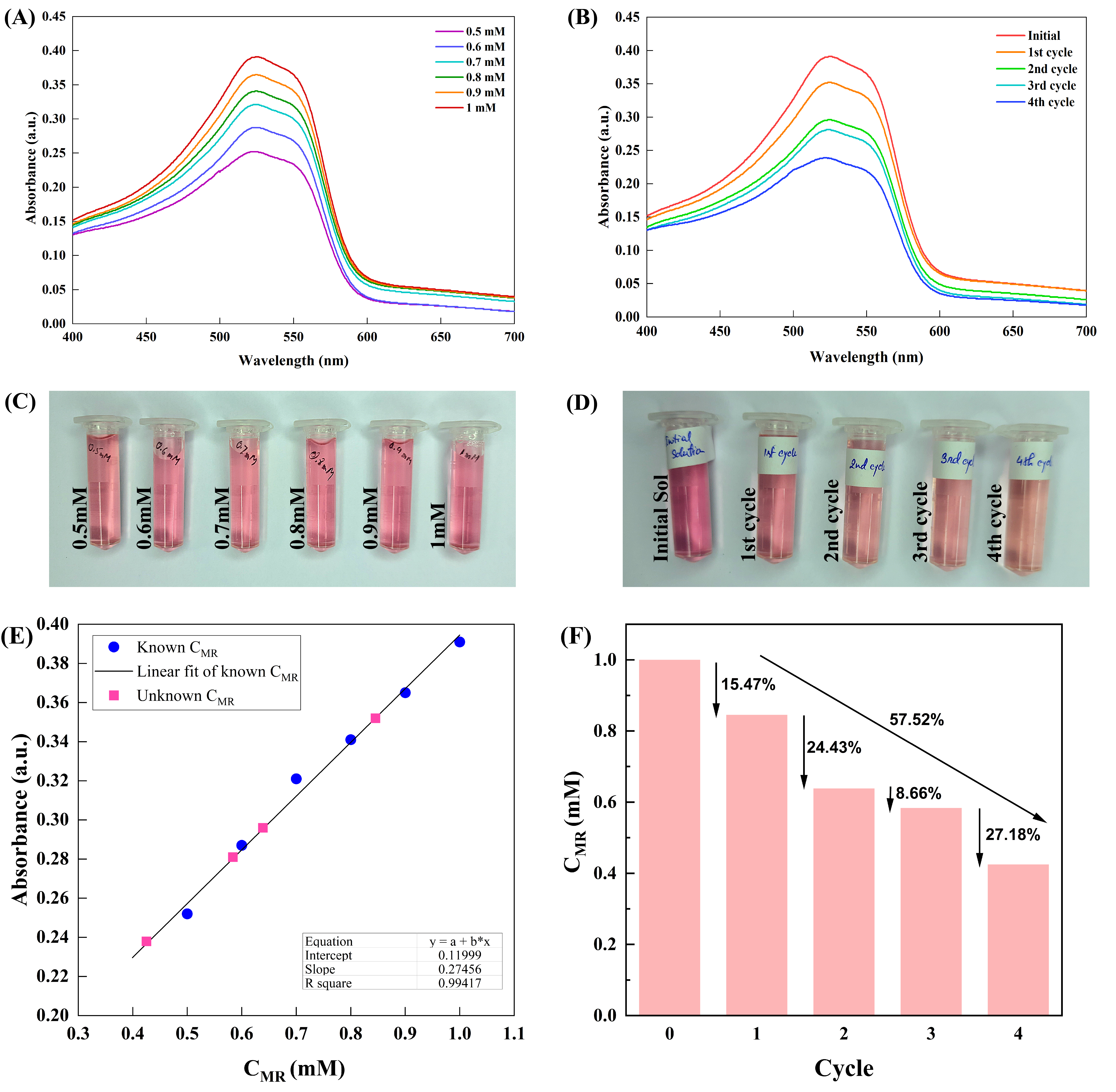} 
    \caption{(A) Variation of UV-vis spectra of known concentration of methyl red dye (C\textsubscript{MR}). (B) Variation of UV-vis spectra of the initial solution and permeates after each cycle. (C) Visual image of dye solutions of known concentrations. (D) Visual image of dye solution of initial and collected permeate samples. (E) Variation in absorbance with C\textsubscript{MR} is calculated from known samples; therefore, the concentration of unknown samples can be found from the linear fit. (F) Change in concentration of methyl red after each cycle.}
    \label{Figure 6}
\end{figure}

To further evaluate the membrane’s separation efficiency, the collected permeate is reintroduced into the microchannel for additional filtration cycles. The UV-visible spectra of the initial solution and permeates after each cycle illustrate the progressive decrease in dye concentration (C\textsubscript{MR}) in permeates, as shown in the plots attached below (Fig. \ref{Figure 6}). The intensity of the methyl red peak at 525~nm decreases with decreasing concentration of methyl red (C\textsubscript{MR}) in permeates. This result indicates that the membrane consistently reduces the amount of dye in the permeate through multiple filtration cycles. From the calibration curve obtained from the known sample data, the C\textsubscript{MR} values of the permeate collected after each cycle can be calculated (Fig. \ref{Figure 6}E).\\

The accuracy of the UV-visible spectra measurements was ensured by using a calibrated spectrophotometer and employing a standardized protocol for sample preparation and measurement. Known concentrations of methyl red dye were used to generate a calibration curve (Fig. \ref{Figure 6}E), which served as the reference for determining the concentrations of unknown samples. This approach minimized errors due to instrument variability and ensured reliable quantification. Furthermore, the spectrophotometer was calibrated prior to each measurement session using standard solutions to maintain consistency and reliability.\\

The filtration process through the membrane demonstrated consistency, with multiple experiments confirming the reliable occurrence of filtration under identical conditions. However, the extent of concentration changes after each filtration cycle varied, influenced by factors such as the membrane's pore size, thickness, and other experimental parameters affecting filtration efficiency. To ensure the reliability of the measurements, experiments were repeated under identical conditions, and the results were found to be reproducible with negligible variation in absorbance values.

\subsection{Microplastic Separation}
\begin{figure}[H]
    \centering
    \includegraphics[width=0.9\textwidth]{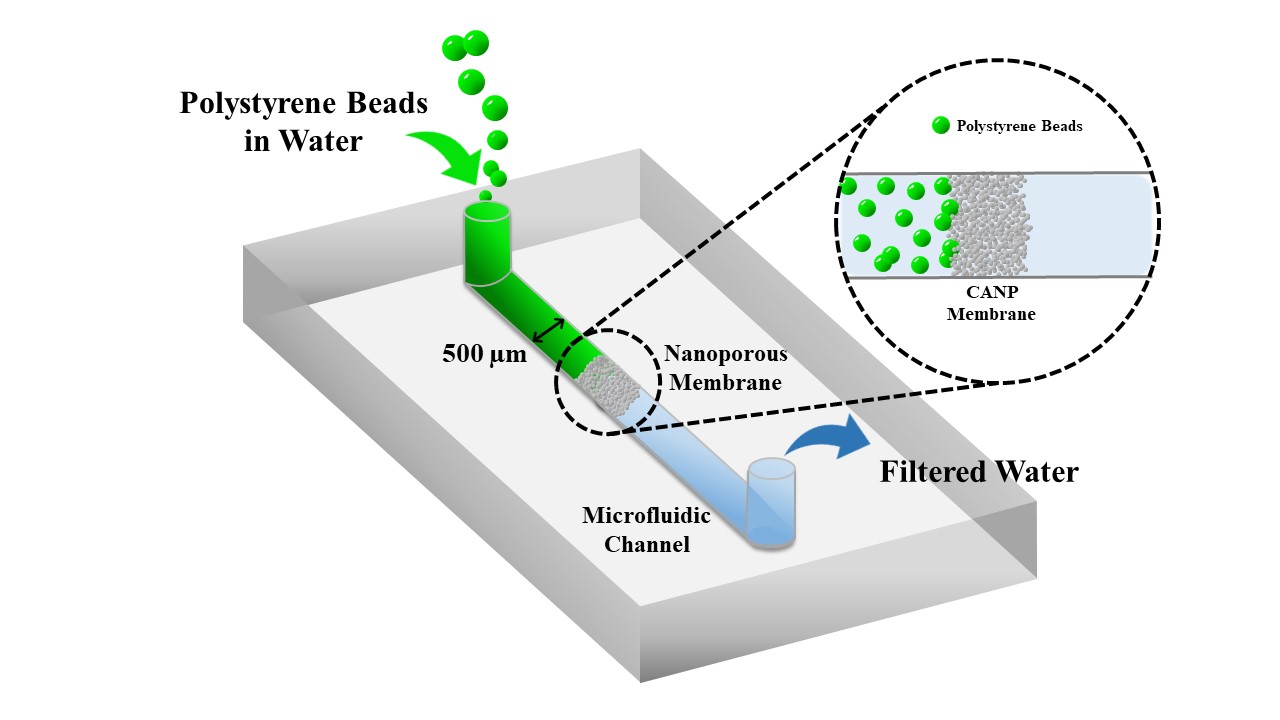} 
    \caption{Filtration of fluorescent polystyrene beads using cellulose acetate nanoporous membrane within microchannel.}
    \label{Figure 7}
\end{figure}

A demonstration was performed to illustrate the water purification capabilities of the synthesized nanoporous cellulose acetate membrane inside the microfluidic membrane with polystyrene beads as a microplastic model. A suspension of 10~$/mu$L of 2~µm green fluorescent polystyrene beads was prepared in 10~mL of water and introduced into the microfluidic channel. The solution was directed across the cellulose acetate membrane at a constant rate of 0.01~mL/min (Fig. \ref{Figure 7}).\\

\begin{figure}[H]
    \centering
    \includegraphics[width=0.9\textwidth]{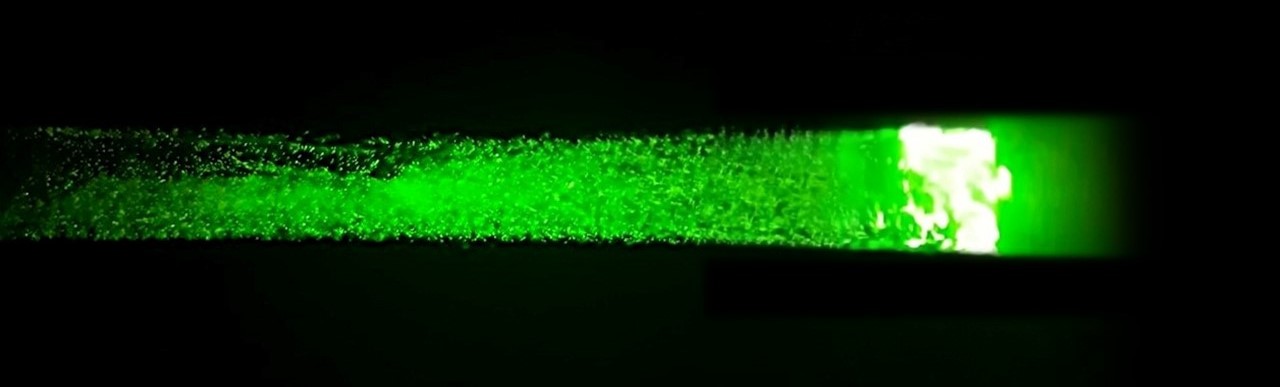} 
    \caption{Microscopic image of fluorescent beads trapped within the fabricated membrane.}
    \label{Figure 8}
\end{figure}

The subsequent microscopic image (Fig. \ref{Figure 8}) clearly shows the accumulation of green fluorescent beads near the membrane interface, indicating their retention within the membrane structure. This image indicates that the polystyrene beads were effectively entrapped within the pores of the membrane, preventing their passage. The filtration efficacy was confirmed by evaluating the filtered water, which revealed a minor quantity of polystyrene beads, representing the membrane's ability to remove microplastics from water. These findings demonstrate the membrane's suitability for application in microplastic filtration and water purification.

\subsection{Membrane Durability and Performance Post-Filtration}
The durability of the fabricated nanoporous cellulose acetate (CA) membrane was assessed during and after the filtration experiments to identify any signs of wear, degradation, or clogging. In the initial filtration cycles, the membrane exhibited consistent performance, with stable flow rates and no significant changes in pressure drop.\\

\begin{figure}[b!]
    \centering
    \includegraphics[width=0.8\textwidth]{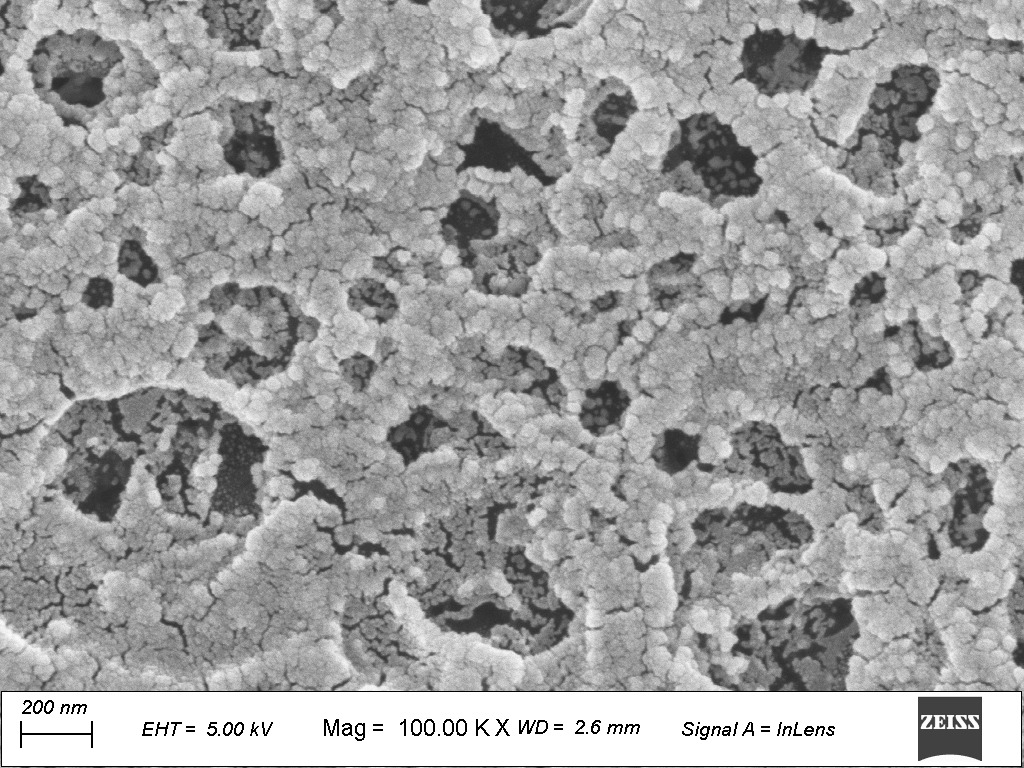} 
    \caption{FESEM image of the cellulose acetate (CA) membrane after the 5th filtration cycle, showing slight wear and tear around pore edges with overall structure intact.}
    \label{Figure 9}
\end{figure}

FESEM images of the membrane, taken after the 5th cycle of dye filtration (Fig. \ref{Figure 9}), revealed slight wear and tear on the surface, particularly around the pore edges. This suggests minor mechanical stress or interactions with dye particles during extended use. Despite these observations, the overall membrane structure and pore integrity remained largely intact, resulting in only a negligible impact on its filtration efficiency.\\

Additionally, a slight decrease in flow rate and a marginal increase in pressure drop were observed after extended filtration cycles, indicating minor clogging caused by the accumulation of dye molecules on the membrane surface. A similar trend was observed during microplastic filtration, where the membrane maintained its structural stability and functionality despite minor clogging and surface wear under prolonged usage.\\

To further validate the practical applicability of the fabricated nanoporous cellulose acetate (CA) membrane, future experiments are planned to test its performance with real-world water samples containing complex mixtures of microplastics and dyes together. These samples will be collected from diverse sources, including municipal wastewater, industrial effluents, and natural water bodies, to account for variations in particle size, polymer type, and concentration. Additionally, the long-term durability of the membrane will be assessed by subjecting it to continuous filtration of these complex samples to monitor potential wear, clogging, or performance degradation.\\

\section{Conclusion}
In this study, we report a novel approach to in-situ fabrication of nanoporous cellulose acetate membranes in microfluidic devices. Herein, the coagulation of cellulose acetate nanoparticles (CANPs) in a microfluidic channel creates membranes with adjustable pore size and distribution. By leveraging the benefits of microfluidics, this approach aims to simplify the fabrication process while enhancing the performance and versatility of the resultant membranes. Characteristic evaluation of the membrane's ability to separate dye and microplastic gives promising results in this direction. Consequently, the methyl red dye separation experiment demonstrates the membrane's ability to significantly reduce dye concentrations in permeates throughout multiple filtration cycles. Furthermore, the microplastic filtration test provides confirmative evidence of the membrane's capacity to successfully retain polystyrene beads, suggesting potential applications in water purification.\\

This discovery establishes the framework for future studies on the use of these membranes in biological, environmental, and industrial applications. The multifaceted improvements in incorporating nanoporous membranes into microfluidic devices, such as enhanced control over fluid dynamics and separation efficiency, usher tremendous potential in every related field. Future research could focus on improving membrane characteristics, expanding production capacity, and integrating these systems into practical applications. Overall, our findings demonstrate the advantages of the proposed method in producing high-quality nanoporous membranes and highlight the broader implications for the development of advanced microfluidic systems with enhanced functionality to pave the way for superior filtration and separation technologies.\\

\backmatter

\section*{Acknowledgements}
\addcontentsline{toc}{section}{Acknowledgements}
\phantomsection
\pdfbookmark[1]{Acknowledgements}{ack}

The authors thank the Centre for Nanotechnology IIT Guwahati, the Central Instruments Facility IIT Guwahati, and the Department of Chemical Engineering IIT Guwahati for the various characterization facilities.\\

\bibliography{sn-bibliography}

\end{document}